# Design and operation of a microfabricated phonon spectrometer utilizing superconducting tunnel junctions as phonon transducers

O O Otelaja[1,2], J B Hertzberg[2,3], M Aksit[2] and R D Robinson[2,4]

[1] School of Electrical and Computer Engineering, Cornell University, Ithaca, NY 14853, USA
[2] Department of Materials Science and Engineering, Cornell University, Ithaca, NY 14853, USA
Email: rdr82@cornell.edu.

**Abstract.** In order to fully understand nanoscale heat transport it is necessary to spectrally characterize phonon transmission in nanostructures. Towards this goal we have developed a microfabricated phonon spectrometer. We utilize microfabricated superconducting tunnel junction-based (STJ) phonon transducers for the emission and detection of tunable, non-thermal, and spectrally resolved acoustic phonons, with frequencies ranging from ~100 to ~870 GHz, in silicon microstructures. We show that phonon spectroscopy with STJs offers a spectral resolution of ~15-20 GHz, which is ~20 times better than thermal conductance measurements, for probing nanoscale phonon transport. The STJs are Al-Al$_x$O$_y$-Al tunnel junctions and phonon emission and detection occurs via quasiparticle excitation and decay transitions that occur in the superconducting films. We elaborate on the design geometry and constraints of the spectrometer, the fabrication techniques, and the low-noise instrumentation that are essential for successful application of this technique for nanoscale phonon studies. We discuss the spectral distribution of phonons emitted by an STJ emitter and the efficiency of their detection by an STJ detector. We demonstrate that the phonons propagate ballistically through a silicon microstructure, and that submicron spatial resolution is realizable in a design such as ours. Spectrally resolved measurements of phonon transport in nanoscale structures and nanomaterials will further the engineering and exploitation of phonons, and thus have important ramifications for nanoscale thermal transport as well as the burgeoning field of nanophononics.

---

[3] Current address: Joint Quantum Institute, University of Maryland, College Park, MD 20742

[4] Author to whom correspondence should be addressed

**Contents**

## 1. Introduction

*1.1. Importance of nanoscale phonon spectroscopy*

One of the grand challenges of nanoscience is to develop experimental tools to understand the fundamental science of heat flow at the nanoscale [1, 2]. In insulators and dielectrics, acoustic phonons are the dominant heat carriers [3], [4]. In nanostructures, as the sample's dimension or surface morphology becomes comparable to phonon characteristic lengths — wavelength, mean free path, and coherence length — the interactions of phonons with these structural features lead to regimes of phonon propagation in which the effect of confinement, scattering, and/or interference of phonons dominates heat transport [5], [6]. To probe these nanoscale effects on phonon transport, one needs a measurement technique that can precisely distinguish wavelength (or frequency) and position of the phonon modes. Previous studies have investigated the effects of nanoscale geometries on thermal transport using Joule-heated metal films on suspended structures [7], [8], [9], [10], [11], but because a thermal conductance measurement employs a broad spectral distribution of phonons, the frequency dependence of the phonon transport in such measurements is difficult to distinguish. Therefore, there is a strong need for a nanoscale technique that will spectroscopically measure phonon transport at hypersonic (>1 GHz) frequencies — particularly at frequencies above 100 GHz which are most relevant to heat flow [12]. Such a technique will be apt for the development of the burgeoning field of nanophononics [13].

An ability to fully understand the propagation of phonons will inform the engineering and exploitation of nanostructures and nanomaterials. For instance, through careful phonon engineering the realization of more efficient thermoelectric materials and microelectronic coolers will be feasible [10, 14, 15]. Such phonon engineering strategies have been recently demonstrated with silicon phononic crystal structures, which displayed a reduction in phonon thermal conductivity in comparison to bulk crystals [16] [17]; however, the exact mechanism and frequency dependence of this reduction is not completely understood because diagnostic tools for nanoscale phonon spectroscopy were not available.

In this paper we describe a new tool for nanoscale phonon spectroscopy using microfabricated superconducting tunnel junctions (STJs) – we detail its design and principle of operation, the fabrication techniques and challenges, the instrumentation and measurement procedures, and the results of selected phonon transport measurements. Phonon spectroscopy with STJs uses a narrow, non-thermal, and tunable frequency distribution of acoustic phonons to probe the phonon transport through nanostructures. STJ-based phonon spectroscopy has previously been performed extensively in macroscale samples by only a few research groups [18, 19] [20] [21]. However, with the development in recent years of advanced micro/nanofabrication techniques, the phonon spectrometer can now be fabricated at the microscale and offer exceptional spatial resolution. The microfabricated phonon spectrometer has the advantage of probing nanoscale effects such as phonon confinement [3], end-coupling diffraction [22], and surface scattering [23], with submicron spatial resolution. We have recently demonstrated a prototype microfabricated spectrometer for emission and detection of non-equilibrium phonons with frequencies ranging from 0 to ~200 GHz [24], and have now tuned the phonon source (emitter) to emit phonons with frequency ranging from 0 to ~870 GHz. The spectrometer comprises a pair of aluminum-aluminum oxide-aluminum (Al-Al$_x$O$_y$-Al) superconducting tunnel junctions serving as phonon emitter and phonon detector on opposite sides of a silicon microstructure. The spectrometer measures the rate of phonons that propagate ballistically through the microstructure. Here we discuss in full detail the design, fabrication steps, required characterization, electronics, and measurement techniques involved in successfully realizing phonon spectroscopy with microscale STJ phonon transducers.

*1.2. Spectrometer design*

The device design for each spectrometer consists of two STJ phonon transducers — one emitter and one detector — attached on opposite sides of a mesa that is monolithically etched on a silicon substrate (See figure 1a). The mesas, which are ~0.8 μm high and have widths ranging from 7 to 15 μm, allow for the isolation of a ballistic path for phonon propagation. The devices are fabricated on a 525 μm thick silicon (100) wafer and the mesa sidewalls are on the Si (111) plane. (Because the mean free path of phonons at our experimental temperature and frequencies is >>1 mm [25], the detected phonons will also include phonons that backscatter from the bottom of the substrate.) The ballistic path along the <110> direction between emitter and detector may be blocked by etching a trench into the mesa in order to determine this contribution of backscattered phonons [24]. This phonon transport measurement platform also enables the monolithic integration of nanostructures into the mesa. Microfabrication methods make the experiments very scalable – spectrometers are fabricated in lots of 100 on 100 mm Si wafers.  Each 4.5 mm square chip contains up to 6 spectrometers, as shown in figure 1b.

The phonon emitter is a single Al-Al$_x$O$_y$-Al tunnel junction with the majority of the junction area lying on the sidewall of the mesa.  The aluminum films are designed to be thin enough (<100 nm) to ensure that the decay length of the phonons is greater than the film thickness in order to minimize phonon reabsorption (emitted phonons breaking quasiparticles within the emitter film) [26].  As will be described, we isolate narrow bands of phonon energy by modulation of the emitter voltage.  Emitter junction resistance should therefore be made low enough to maximize the amount of current (and therefore phonon signal) flowing at a given modulation amplitude, while the resistance must also be large enough to inhibit overinjection of electrons through the tunnel barrier into the Al film.  Such overinjection may locally suppress the superconducting gap and thereby degrade energy resolution [27].  Residual inhomogeneities in the gap are inherent in the film and may be assessed from an I-V curve of the junction.  Typically we use emitters having junction resistances from ~800 to ~5000 Ω, and we observe an inhomogeneity of about ~60 to ~80 μeV (~15 to ~20 GHz) which represents the upper limit of our energy resolution.

The detector is designed to have a double-junction (SQUID geometry) with a 'hot electron finger' extending onto the mesa sidewall to capture the incident phonons.  The actual detector junctions lie on the (100) plane of the silicon substrate.  The SQUID geometry enables the suppression of Josephson current via the application of a magnetic field.  With the Josephson current suppressed, we can readily distinguish an incident phonon flux as an increase in the 'subgap' tunnel current due to the incident phonons breaking Cooper pairs in the Al.  Incident phonons of energy $\geq 2\Delta_d$ break Cooper pairs in the aluminum film of the detector finger, and the excited quasiparticles diffuse to the junctions and tunnel through the oxide barrier.  Some of the detectors have quasiparticle traps made from thin Au films.  These traps are designed to ensure that excess quasiparticle energies do not reach the junction, and they also prevent back tunneling of quasiparticles [28].  When we change the length of the detector fingers, $L_f$, from 10 μm to 20 μm (see figure 1a), we found no discernible difference in the detected phonon transmission signal levels.  We conclude from this that the quasiparticle diffusion length is much longer than the finger length, a conclusion that agrees with diffusion lengths reported in the literature [29].  The tunnel junction therefore faithfully measures the rate of phonon arrival at the tip of the finger several microns distant.  Forming the double-junctions on the flat Si (100) plane reduces their asymmetry , therefore facilitating Josephson-current suppression, simplifies fabrication, and offers great flexibility in spatial resolution to be achieved merely by changing the width and position of the finger [24].

## 2.   Principles of operation

*2.1. Phonon emission with STJ*

Phonon emission in STJs occurs via the excitation and decay of quasiparticles (single electrons) in superconducting films.  As depicted in figures 2a and 2b, when the emitter STJ is biased above the superconducting gap ($2\Delta_e$) such that $V_e \geq 2\Delta_e/$e ($V_e = I_e R_n$, where $I_e$, $V_e$, and $R_n$ are the current through, voltage across, and normal state tunneling resistance of the emitter junction respectively), the Cooper pairs (paired electrons) in the first aluminum film break apart and quasiparticles tunnel through the oxide barrier into excited energy states ranging from $\Delta_e$ to e$V_e - \Delta_e$ (all  energies referenced to the Fermi level in the aluminum film at the opposing side of the junction) [18, 19].  The Al emitters in our experiments have $2\Delta_e$ of ~400 μeV at a temperature of ~0.3 K.  These excited quasiparticles rapidly decay towards the edge of the superconducting gap, emitting phonons as they decay. Due to the singularity in the density of states at the gap edge, this 'relaxation' process typically requires only one or two decay steps before the quasiparticle energy is reduced to $\Delta_e$. This process thus emits a broad distribution of phonons of energies ranging from 0 to $eV_e - 2\Delta_e$. The phonons are incoherent and to a first approximation will have both random polarization and random direction due to elastic scattering of the tunneled electrons within the Al film. The shape of this 'relaxation' phonon distribution includes a sharp cutoff at energy $eV_e - 2\Delta_e$, thus allowing a small modulation of $V_e$ to isolate a narrow portion of the spectrum that is sharply peaked at energy $eV_e - 2\Delta_e$. Subsequent recombination of the quasiparticles into Cooper pairs lead to the emission of recombination phonons of energy $2\Delta_e$. The average relaxation $\tau_{rel}$ and recombination $\tau_{rec}$ times are on the order of ~1 ns and 30 µs respectively [30]. When the STJ emitters are attached to one end of a microstructure, relaxation and recombination phonons — both longitudinal and transverse polarizations — are emitted and ballistically propagate through the microstructure; however, only the relaxation phonons are controlled by modulation techniques for spectroscopic studies.

*2.2. Modeling the phonon emission spectrum*
When considering spectral precision of a phonon source, a convenient figure of merit is the ratio $P_{peak}/P_{tot}$ of phonon power near the peak of the distribution, to the total power in the measurement. For instance, a thermal conductance measurement at temperature $T$ employs a Planck's distribution of phonon modes, having power spectral density $P(\omega)d\omega \sim \frac{\hbar\omega^3 d\omega}{2\pi^2 v^3}/\left(e^{\hbar\omega/k_B T} - 1\right)$, where $\omega$ is the angular frequency of the phonon and $v$ is the phonon speed. This distribution is peaked at the so-called 'dominant phonon frequency' $\omega_{dom} = 2.78 k_B T/\hbar$, but the distribution is quite broad and therefore a slice of spectrum within $\delta\omega$ around the peak contains only a small fraction of the total power. For instance, if we wish to interrogate a spectral feature at 400 GHz with 20 GHz precision, a Planck distribution at $T = 6.9$ K offers $\omega_{dom}/2\pi = 400$ GHz but contains only 3% of its power within +/-10 GHz of this peak.

To model the phonon emission profile of the modulated STJ phonon spectrum and to estimate $P_{peak}/P_{tot}$ of this distribution, we must carefully consider the non-equilibrium electron-phonon interactions within the superconducting film in the emitter STJ. These include phonon attenuation due to Cooper-pair breakage [26, 30, 31] and acoustic-mismatch transmission across the Al/Si boundary [32] as well as quasiparticle diffusion and reemission of absorbed phonons. The total emitted phonon power resulting from the quasiparticle relaxation process will comprise the phonons emitted in first-step relaxation, plus any emitted in second-step relaxation, minus the fraction reabsorbed by Cooper pair breakage within the aluminum, plus the power that is reemitted following this reabsorption processes.

For voltages $V_e$ not greatly exceeding $2\Delta_e$, nearly all injected quasiparticles decay to energy $\Delta_e$, so that first-step relaxation dominates, the entire modulated spectral power falls at frequency $(eV_e - 2\Delta_e)/h$, and if we neglect the effect of reabsorption then $\frac{P_{peak}}{P_{tot}} \cong 1$ [19]. For higher bias voltages, a fraction of the quasiparticles will relax first to intermediate energies before undergoing secondary relaxation to the band edge energy $\Delta_e$. The precise distribution of generated phonon energies may be found by convolution integral of the quasiparticle injection rates, densities of states and decay rates [30, 33]. For simplicity, we will assume that the phonon density of states in the Al follows a Debye model, and adopt an approximate model of phonon production rate, presented by Eisenmenger et al., based on the spectrum of phonons emitted by first-step relaxation of electrons injected across a normal-state tunnel junction of resistance $R_n$ at $T = 0$ [30, 34].

$$\dot{N}_e(\omega)d\omega = \frac{3}{e^2 R_n}\left(1 - \left(\frac{\hbar\omega}{eV_e}\right)^2\right)d\omega \qquad (1)$$

This rate of phonon production per unit bandwidth $\dot{N}_e(\omega)$ is shown in figure 2c, and extends from $\omega = 0$ to a sharp cutoff at $\omega = (eV_e - 2\Delta_e)/\hbar$. This is a good approximation for voltages $eV_e \gg 2\Delta_e$ [30, 34]. It is evident from the shape of this distribution that a portion of the differential phonon power is produced at the peak $\hbar\omega = (eV_e - 2\Delta_e)$ while the remainder is produced at energies broadly distributed over the range 0 to $eV_e - 2\Delta_e$. The power spectral density $P(\omega)d\omega$ may be found from equation (1) as $P(\omega)d\omega = \hbar\omega \dot{N}_e(\omega)d\omega$. Second-step relaxation may add up to 25% additional phonon power, mostly at frequencies well below the cutoff at $(eV_e - 2\Delta_e)/\hbar$ [30].

We must also consider reabsorption and reemission of phonon energy. Attenuation of the phonon population within the superconductor will occur as phonons of energy $\hbar\omega > 2\Delta_e$ break Cooper pairs, creating fresh quasiparticles. The probability that a phonon will survive traveling a distance $r$ within the aluminum is $e^{-r/\Lambda_{ph}}$, the mean absorption length of $\Lambda_{ph}(\omega)$ being dependent on phonon energy $\hbar\omega$ and band gap energy $\Delta_e$ [30]. If we treat the phonons as point-particles traveling ballistically within the Al, then the probability of a phonon generated at a distance $z$ from the Al/Si interface and traveling at an angle $\theta$ to the normal, to escape into the Si before reabsorption is [30, 35]

$$e^{-z/(\Lambda_{ph}\cos\theta)}T_{AlSi}(\theta). \qquad (2)$$

Here $T_{AlSi}(\theta)$ is an acoustic-mismatch transmission factor for wave transmission from Al into Si. The films of some of our emitter STJs have lower and upper layer thicknesses of ~20 nm and ~79 nm respectively on the mesa sidewall (as determined by profilometry measurement and adjusted for sidewall angle). For simplicity, we treat all phonons as being generated within the lower layer at a spatially uniform rate. We assume the phonons' velocities are distributed uniformly in all directions, and that those entering the top layer may reflect from the Al/vacuum boundary, reenter the lower layer, and reach the Al/Si boundary. For phonons to emerge and travel directly across the mesa towards the detector (an angle ~35.3 degrees to the sidewall normal), we estimate the refraction angle within the Al using Snell's law, assuming average wave speeds $\overline{v_{Al}} = 4.4 \times 10^3$ in Al and $\overline{v_{Si}} = 6.6 \times 10^3$ m/s in Si, to be $\theta \sim 22.7°$. From reported values of the acoustic impedances of Al and Si, we estimate $T_{AlSi}$ to be > 0.9 for such an angle and to be frequency-independent [6, 30]. Kaplan et al. have calculated values for phonon decay time in Al as a function of phonon energy $\hbar\omega$ and bandgap energy $\Delta_e$ [26]. We multiply these by $\overline{v_{Al}}$ to find $\Lambda_{ph}(100\text{ GHz}) \cong 1.04$ μm, $\Lambda_{ph}(400\text{ GHz}) \cong 0.38$ μm and $\Lambda_{ph}(700\text{ GHz}) \cong 0.22$ μm. While these values are greater than some reported experimental values of $\Lambda_{ph}$ in Al at energy $\hbar\omega = 2\Delta_e$, they are comparable to measured values of normal-state acoustic attenuation corrected to the[30-32, 35-37] superconducting state. Averaging equation (2) over our full Al layer thicknesses, we estimate that in the direction pointing out towards the detector, ~90% of phonons at $\omega/2\pi = 100$ GHz will escape into the Si, ~78% at $\omega/2\pi = 400$ GHz and ~68% at $\omega/2\pi = 700$ GHz. We use these attenuation factors to modify the spectrum in equation (1), as shown in figure 2c.

To find the total rate of absorbed phonons, we must average equation (2) over all depths and angles. At large values of $\theta$ we note that $T_{AlSi}(\theta)$ will be <<1, regardless of phonon frequency, and for angles above about 45 degrees, $T_{AlSi}(\theta)$ will be zero due to total internal reflection within the Al [30, 32]. Transmission coefficients $\langle T_{AlSi}\rangle$ averaged over all angles and phonon polarizations have been calculated by Kaplan, from which we estimate $\langle T_{AlSi}\rangle \sim 0.44$ assuming the three phonon polarizations to be equally populated [32]. Thus at any frequency $\omega > 2\Delta_e/\hbar$, at least 56% of all phonons produced are liable to be reabsorbed within the Al. We can approximate the additional frequency-dependency by multiplying this $\langle T_{AlSi}\rangle$ by the average of equation (2) over the full Al layer thickness and all angles less than the critical angle. Therefore among all phonons at all angles we estimate that ~61% are reabsorbed at $\omega/2\pi = 100$ GHz, ~67% at $\omega/2\pi = 400$ GHz, and ~71% at $\omega/2\pi = 700$ GHz. For each bias voltage $V_e$, we apply these proportions to the spectrum of equation (1) and integrate to find the total reabsorbed power.

By conservation of energy, all of this reabsorbed power must be reemitted. The quasiparticles created in the reabsorption subsequently relax and recombine to emit additional phonons of lower frequency than the ones initially absorbed. We estimate based on typical decay times and on the geometry of our STJ on the mesa sidewall that the quasiparticles do not travel far prior to reemission, so that about 80% of the power is reemitted at the same or nearby location as the original tunneling injection in the Al film on the mesa sidewall.

Taking together first-step relaxation, second-step relaxation (constituting up to ~25% of the total relaxation phonon power), attenuation, and reabsorbed/reemitted power, we find that for typical $V_e$ values of up to a few mV, the total modulated power $P_{tot}$ emitted from the emitter STJ is roughly proportional to the modulated emitter current $\delta I_e$. The power emitted due to recombination on the other hand (see figure 2a) should remain fixed as $V_e$ is varied, and for large $V_e$ we take this to be a negligibly small fraction of the total power. Therefore the total emitted differential phonon rate is $\sim \delta I_e/e$.

To find $P_{peak}/P_{tot}$ at a given peak frequency $\omega_{peak}$, we take $P(\omega_{peak})\delta\omega$ from equation (1), for a given peak width $\delta\omega$ (e.g. $\delta\omega/2\pi = 20$ GHz), attenuate this quantity according to equation (2) as described above, and divide by the total power $P_{tot}$ found as described above at $V_e = (\hbar\omega_{peak} + 2\Delta_e)/e$. The result of this calculation for our typical emitter film thicknesses appears in figure 2d. For a peak width $\delta\omega/2\pi = 20$ GHz, at a peak frequency of $\omega/2\pi = 100$ GHz, $P_{peak}/P_{tot}$ is ~50%. This diminishes to ~32% at peak $\omega/2\pi = 400$ GHz, and further at higher peak frequencies. As shown in figure 2d, the values of $P_{peak}/P_{tot}$ from the STJ-emitted phonon spectrum compare very favorably to a Planck distribution, exceeding it by more than an order of magnitude for $\omega/2\pi > 300$ GHz. This analysis demonstrates that aluminum STJs made of films a few tens of nm thick will emit narrow spectral distributions of acoustic phonons into Si at frequencies up to several hundred GHz.

Phonon emission from aluminum STJs has been reported elsewhere at frequencies up to ~2 THz, but $P_{peak}/P_{tot}$ is likely to be very small at such a peak frequency even if the films are made very thin [38]. The wavelength in Al at 700 GHz is ~6 nm while the granularity in the Al film and the roughness at the Al/Si interface are most likely a few nm; hence, for $\omega/2\pi$ above ~700 GHz, we expect to see the spectrum further modified by the effects of elastic scattering of phonons within the Al film[35], inelastic phonon scattering at the Al/Si boundary [39] and modification of phonon spectra due to excess injected quasiparticle population in the Al film [27]. All such effects are liable to become more severe as $V_e$ and $\omega$ are increased.

*2.3. Phonon detection with STJ*

The phonons incident on the detector are registered as an increase in the tunnel current through the detector junctions. The STJ detector is biased below its superconducting gap with voltage $V_d < 2\Delta_d/e$ (figure 2b). Phonons incident on the detector finger with energy greater than or equal to $2\Delta_d$ will break Cooper pairs in the detector films, and the quasiparticles will diffuse until a portion reaches the detector junction and tunnel through. The STJ detectors are made from aluminum films with superconducting gap $2\Delta_d$ ~360 μeV (corresponding to ~90 GHz), and in essence these detectors act as high pass filters of acoustic phonons with cut-off frequency ~90 GHz. A lock-in detector selects only the modulated portion $\delta I_d$ of the detector current, corresponding to the modulated emitter phonons that strike the detector. The phonon spectrum therefore comprises phonons of frequencies between ~90 GHz and $(eV_e - 2\Delta_e)/h$, with a sharp peak at frequency $(eV_e - 2\Delta_e)/h$. Because the modulated emitter phonon power is proportional to $\delta I_e$, the measured differential transfer function $\delta I_d/\delta I_e$ tells us the fraction of this spectrum that is transmitted from emitter through the sample to the detector.

*2.4. Modeling the detector behavior*

We may use quasiparticle-phonon interactions to model and quantify the phonon detector behavior.

For a differential rate $\dot{n}_{ph,d}$ of phonons of frequency $\omega$ striking the detector finger, we expect the average differential rate of phonon-induced quasiparticle generation $\dot{n}_{QP,ph}$ to be

$$\begin{aligned}
\dot{n}_{QP,ph} &= 0 & \text{for } \hbar\omega < 2\Delta_d \\
\dot{n}_{QP,ph} &= T_{SiAl} \cdot \alpha_{abs}(\omega) \cdot 2\dot{n}_{ph,d}(\omega) & \text{for } 2\Delta_d \leq \hbar\omega < 4\Delta_d \\
\dot{n}_{QP,ph} &= T_{SiAl} \cdot \alpha_{abs}(\omega) \cdot 2\left(\frac{\hbar\omega}{2\Delta_d} - 1\right)\dot{n}_{ph,d}(\omega) & \text{for } \hbar\omega \geq 4\Delta_d
\end{aligned} \quad (3)$$

In equations (3), $T_{SiAl}$ is the acoustic transmission factor for phonons transiting from Si into Al, which we estimate from acoustic impedances to be >0.9 over all incidence angles [32]. The fraction of phonons $\alpha_{abs}(\omega)$ absorbed in the finger will be approximately $\alpha_{abs}(\omega) = 1 - e^{2d/\Lambda_{ph}(\omega)}$. In our detector fingers, the thickness $d$ in the direction of phonon incidence is 140 to 205 nm, thus we expect $\alpha_{abs}(\omega)$ to equal at least 0.2 for $\omega/2\pi = 100$ GHz, and at least 0.8 for $\omega/2\pi = 700$ GHz. In our devices, the diminishing fraction $P_{peak}/P_{tot}$ as peak frequency is increased (figure 2d) motivates us to treat $\alpha_{abs}$ as independent of peak frequency and having value $\alpha_{abs} \sim 0.25$. In the signal of a typical spectrometer transmitting through bulk Si, we see a modulated signal that is consistent with this assumption and with the detector response behavior of equations (3).

To find $\dot{n}_{QP,ph}$ and thereby the phonon arrival rate $\dot{n}_{ph,d}$ from the measured differential detector tunnel current $\delta I_d$, we must account for quasiparticle loss processes in the detector. The primary loss process comprises diffusion of the quasiparticles into the attached wiring leads, followed by recombination into Cooper pairs [30, 40, 41]. Using conventional theories of tunneling rate and quasiparticle recombination, we may express a nondimensional efficiency factor $\{\text{Eff}\} = \delta I/e\dot{n}_{QP,ph}$ for each detector (see Appendix B) [30, 39, 41]:

$$\{\text{Eff}\} = \sqrt{\frac{\tau_{rec}}{D}} \cdot \frac{1.15}{2e^2 R_n N_0 W_{tr} d_{tr}} \quad (4)$$

where $R_n$ is the normal-state tunneling resistance of the junction, $N_0$ is the normal density of states at the Fermi level ($1.75 \times 10^{10} \cdot \mu m^{-3} eV^{-1}$ in Al) [30], and $W_{tr}$ and $d_{tr}$ are respectively the average total width and thickness of the wiring trace connected to the detector STJ. The factor $D \cong 20$ cm$^2$/s is the diffusion constant for quasiparticles in Al, and $\tau_{rec} \sim 30\mu s$ is the average quasiparticle recombination time in Al at a temperature of 0.3K [30, 42, 43] [29, 40]. In our detectors {Eff} is typically ~ 0.1.

## 3. Fabrication techniques and challenges

Figure 3a illustrates the step-by-step fabrication of the mesas and transducers. The mesas are formed by a shallow depth anisotropic etching of silicon using KOH (50% KOH, 48 °C, 4 min.) with a low-stress silicon nitride etch mask. We found that standard RCA cleaning of the wafers prior to etching is crucial to obtaining smooth surfaces. The smoothness of the (100) and (111) planes is necessary to enable deposition of continuous Al films, and to minimize phonon scattering from rough surfaces. Simultaneous magnetic stirring and ultrasonication during the KOH etch helps to improve the smoothness of the etched surfaces. The trenches are simultaneously formed, where needed, on the mesas. Neither hydrochloric acid nor surfactants was added.

We fabricate the emitter tunnel junctions on the sidewall of the mesa using double-angle evaporation as shown in figure 3b. A bilayer of S1818 photoresist (Rohm and Haas Inc.) and LOR lift-off resist (Microchem Inc.) is spun onto the fabricated mesas and the emitter geometry, wiring trace and bond pads are photolithographically patterned into the resist. The depth of field of our photolithography tool (±2.42 µm) limits the range of mesa heights and resists thickness used to form the junctions. The patterned resist is developed in AZ MIF 300 (AZ Electronic Materials) for ~60 seconds until a "Dolan Photoresist Bridge" is formed with sufficient undercut (see figure 3a and 3b) [44]. The surfaces must be cleaned with Argon or oxygen plasma prior to evaporation to prevent poor aluminum film adhesion and aging of the tunnel junctions formed [45],[46].

As shown in figure 3a, arrays of detector and emitter STJs are patterned and deposited. The film thicknesses in the emitter are made only a few tens of nm (lower layer is ~20 nm thick and upper layer is ~38 to 80 nm thick on the mesa sidewall), to enable phonons to escape into the Si without reabsorption, whereas the detector film thicknesses are made several hundred nm thick to maximize the absorption of incident phonons. We utilize a two-step electron-beam angle evaporation interspersed with a static oxidation procedure to form the aluminum tunnel junctions. The overlap area of the tunnel junctions is dependent on the angles at which the evaporation is done (figure 3b). Assuming that the height of the bridge or thickness of the LOR layer is $d$ μm and the width of the bridge is $W_1$ μm, the overlap width is $d \cdot (tan\theta_1 + tan\theta_2) - W_1$, where $\theta_1$ and $\theta_2$ are the two deposition angles measured from the normal to the substrate surface (see figure 3b). The width of the wiring traces, $W_2$, should be made wide enough that the double angle evaporation forms a single overlapping metal trace (figure 3c). We found that the best quality films were obtained at evaporation rates ~4.5 - 5 Å/s. The films were evaporated at base pressures ranging from 2 x 10$^{-7}$ Torr to 1.2 x 10$^{-6}$ Torr. The base pressures were sometimes lowered further by the initial evaporation of 50-100 nm Al in the chamber. The evaporated aluminum acts as a getter for particles in the chamber. The tunnel barrier is formed by static oxidation in between the two Al deposition steps, with an exposure parameter defined as $exposure\ (Pa-secs) = pressure\ (Pascal) \times time(seconds)$ [47]. The emitter tunnel barrier was grown in 3 Torr of oxygen for 60 minutes resulting in emitter resistances ~1.5 kΩ. The detector tunnel barrier was grown at 300 mTorr for ~70 minutes resulting in resistances of ~200 Ω. Figure 4a shows the exposure parameter plotted against the area-specific resistances of the tunnel junctions. This guide can be used to estimate the exposure parameters that will produce emitters or detectors with desired junction resistances. The plot was fitted to a power law ($f(x) = x^{0.59}$) with an adjusted-$R^2$ value of 0.997. The area of each junction was calculated from scanning electron micrograph (SEM) inspection and can be estimated prior to fabrication based on the overlap area calculations discussed above. The post-evaporation processing includes metal lift-off, dicing of the wafer into 4.5 sq. mm chips, and the evaporation of ~500 nm thick silver on the backside. Silver has been shown to be a good absorber of phonons [25]; hence, the addition of silver reduces the backscattered signals from the bottom of the chip. The junctions are very sensitive to static discharge, and therefore, proper grounding is essential at all times.

The base pressure at which the Al films are deposited is important as it may affect their room-temperature resistivity, which in turn affects their critical temperature $T_c$ and superconducting gap. Such variations in the superconducting gaps of aluminum films with respect to their room-temperature resistivity have been reported to be due to oxygen doping [48], [49]. In figure 4b, we show the dependence of the room-temperature resistivity of thin aluminum films of identical dimensions on increased oxygen partial pressure during evaporation. The dimensions of the films were patterned by photolithography and the first film was evaporated at a base pressure of 0.44 μTorr. By increasing the base pressure due to the continuous flow of oxygen into the chamber, we show that the resistivity of the films varies with base pressure. The typical transition temperature, $T_c$, for films evaporated at base pressures of 27 μTorr and 35 μTorr was measured to be 1.75 K and 1.89 K respectively (increased $T_c$ compared to a pure Al film with $T_c$~1.12 K).

## 4. Instrumentation, measurement technique, and characterization of spectrometer

### 4.1. Low temperature apparatus

The apparatus for the low temperature phonon transport experiments includes a He-3-cryostat with a custom designed sample stage immersed in a liquid helium Dewar. The fridge wiring consists of twisted pair lines with room temperature pi-filters (Tusonix 4701 EMI) enclosed in brass block Faraday cage, allowing up to ~90 dB attenuation at frequencies > 100 MHz. The cold stage filters are 'tapeworm' type low-pass filters [28], but the extent of cold stage filtering is limited by the space in our vacuum can. The fridge is cooled down to a base temperature of 0.3 K and the sample is held in vacuum. The thermometer at the He-3 stage is a Cernox™ RTD (Lakeshore Cryotronics) and a silicon diode thermometer (DT-470-SD-12A, Lakeshore Cryotronics) monitors temperature at the 1 K–pot. Attempts are made to minimize the coupling of noise from the thermometry wiring into the measurement wiring. Metal film resistors are used in all bias networks, as this type of resistor is known to exhibit superior temperature stability and reduced $1/f$ noise. As shown in figure 5a, the chips containing the spectrometers are wire-bonded onto the gold plated copper sample stage. The backside of the chips must be properly anchored to the sample stage by thermalizing with Apiezon N grease or silver paint. A 5000 turn superconducting magnetic coil is attached to the top of the sample box for Josephson current suppression as shown in figure 5b, c, and d. Once the fridge is immersed in the helium Dewar, we ensure proper grounding of the fridge and equipment rack. We place rubber pads underneath the wheels of the Dewar to reduce mechanical vibration.

### 4.2. DC characterization of STJ emitters and detectors

The DC characteristics of emitter and detector tunnel junctions are determined from current-biased current-voltage (I-V) measurements at ~0.3 K. Figure 6a shows how we estimate the superconducting gap from the I-V behavior. From the I-V curves we also calculate the normal state resistance, $R_n$, of the junctions. In figure 6b, we show the current-biased I-V curves in the subgap regime for four SQUID detectors with the current normalized by their normal state resistances for comparison. The red plot shows significant rounding-off which is due to poor filtering on that particular signal line, allowing stray voltage noise to add a random perturbation to the junction voltage. In figure 6c, we show the resistance-normalized I-V curves for four emitters with normal state resistance values of 212 Ω, 935 Ω, 2250 Ω and 5559 Ω. This plot illustrates several possible problems in emitter performance. In the 212 Ω emitter (red plot), we observe 'back bending' of the gap rise step at $V_e = 2\Delta_e/e$. This is a signature of quasiparticle overinjection, which appears consistently in emitter STJs of $R_n$ < 700 Ω, leading to local suppression of the superconducting gap $\Delta_e$ and poor phonon energy resolution. In the 2250 Ω junction (magenta plot), the I-V curve shows a signature of being partially shorted (this could occur either at their formation or during processing) which will add an uncontrolled thermal phonon population to the junction's emission. The black and blue curves indicate a limitation on emitter energy resolution. For an ideal STJ, the 'gap rise' step at $V = 2\Delta/e$ should be infinitely sharp, but in practice, we observe a breadth of ~60-80 μV (~15 to 20 GHz). This behavior most likely indicates that the superconductor's gap $\Delta_e$ varies within the junction by ~60-80 μeV (corresponding to a ~15 to 20 GHz imprecision in emitted phonon frequency).

### 4.3. Josephson current suppression

Josephson current (or supercurrent) in the detector must be suppressed, so that the detector may be voltage-biased and its quasiparticle tunneling current clearly distinguished. To do so, we apply a magnetic field perpendicular to the SQUID loop, using a small superconducting coil mounted as close as possible to the top of the chip to minimize vibration-coupled flux noise. For our coil geometry (see figure 5c), we calculate (using Biot-Savart law) the axial magnetic field to be 1.27 Gauss/mA. The heat load resulting from typical coil current is ≤2 μW. The maximum supercurrent in the SQUID detector junction, assuming perfect symmetry, is given as $I_c(\Phi) = 2I_c(0)\left|(\cos(\frac{\pi\Phi}{\Phi_0}))\right|$, where $\Phi_0$, $\Phi$, and $I_c(0)$ are the flux quantum (2.07 x 10$^{-15}$ Wb), applied flux, and critical current at zero magnetic field respectively [50]. By applying a magnetic flux proportional to $\frac{n\Phi_0}{2}$, where $n$ is an odd integer, the supercurrent should be fully suppressed. We typically employ the minimum effective flux (equivalent to $n = 1$), in order to minimize flux trapping. In practice, we find that the supercurrent is not always fully suppressed, probably due to asymmetry between the two junctions. Figure 7a illustrates our technique for determining the detector bias point for phonon transport studies. The detector voltage is swept in the subgap regime between ~-300 to ~300 µV. At each voltage step, the coil current is swept from 0 to 2 mA and the tunnel current is measured at each step. In the 3D plot in figure 7a, the current measured per the detector bias voltage and per coil current is shown. We set the voltage bias point of the detector to $\sim\Delta_d/e$ (~180 µV) and coil current to ~1 mA, where the minimum critical current is obtained. The measured zero-voltage and zero B-field supercurrent for the detector ($R_n$ = 116 Ω) in figure 7a is ~1.2 µA (z-axis) and is closely predicted by the Ambegaokar-Baratoff expression for $T \sim 0$ K, $I_{c0} = \frac{\pi\Delta}{2eR_n}$ [50]. By applying a magnetic field ~1 Gauss at the bias point, the supercurrent is suppressed to ~1 nA.

The extent to which the supercurrent in the SQUID detectors may be suppressed is dependent on two geometric properties: self-induced flux and junction symmetry. The self-induced flux is proportional to the self-inductance, $L$, of the SQUID loop, which we estimate based on the inductance of a rectangular loop [51]. The more closely identical the two junctions are, the more closely the current flowing through them may be made to cancel. In figure 7b, we plot the ratio of the minimum obtainable critical current to the maximum zero voltage critical current ($I_c(min)/I_{c0}$) versus the parameter $\beta_L = \frac{2LI_c}{\Phi_0}$, the ratio of self-induced flux to the flux quantum. Each symbol in figure 7b represents a unique SQUID design based on the location of the junction and the loop area: Junctions formed on flat surface are represented by solid symbols, while the open symbols represent junctions formed on the sidewall; Loop areas vary from ≤ 2 µm$^2$ (squares), to ~10 µm$^2$ (circles), to ~120 µm$^2$ (triangles), and to ~180 µm$^2$ (diamond). Smaller loop areas and larger junction resistances lead to smaller values of $\beta_L$ and in general to better supercurrent suppression; however, for the SQUID detectors formed on the sidewall, we observe a large variation in suppression for devices with similar $\beta_L$. This is likely due to junction asymmetry. For devices formed on the flat (100) surface, supercurrent suppression is more consistent and exceeds ~3 orders of magnitude for $\beta_L < 2 \cdot 10^{-3}$, indicating more symmetric junction formation. We also note a tradeoff in detector design: while Josephson critical current scales inversely with normal-state tunnel resistance $I_{c0} = \frac{\pi\Delta}{2eR_n}$, detector efficiency (equation (4)) also scales inversely with $R_n$. In practice we find that a loop area of ~2 µm$^2$ and detector resistance $R_n$ ~ 200 to 300 Ω enable both suppression of $I_c$ to levels smaller than thermal quasiparticle tunneling current, as well as detector efficiencies of ~0.1 that permit readily measurable spectrometer signals.

With the supercurrent suppressed, we measured the subgap tunnel current due to thermally excited quasiparticles at detector voltage $V_d = \Delta_d/e$ and at different temperatures (~0.3 to 0.4 K) as shown in figure 7c (red plot). We compare the results to the BCS approximation of the subgap tunnel current for an S-I-S junction (blue plot) [52]. The measurement shows exponential dependence of subgap current on temperature, as predicted by BCS theory. The deviation between the data and prediction may be due to our inability to fully suppress the supercurrent and to possible inaccuracies of our cold stage thermometer at temperatures below ~0.34 K.

*4.4. Modulated phonon transport measurements*

The schematic of our phonon transport experiments is shown in figure 8a. For phonon emission ($V_e \geq 2\Delta_e/e$), the emitter is current biased by applying a DC voltage, $V_b = \frac{V_e}{R_n}R_b$ through bias resistor $R_n$ ~500 kΩ, where $V_e(=I_e R_n)$ is the voltage across the emitter junction and $R_n$ is the normal state resistance of the emitter junction. All the device wiring comprises filtered twisted-pair lines, and shielded coaxial cables are used for all connections. The DC current through the emitter junction is stepped from $I_e = $ ~ 0.35 to 2 µA, which corresponds to emitter voltages $V_e = $ ~0.35 to 2 mV for a junction resistance of $R_n = 1$ kΩ. In addition to the DC current applied to the junction, an AC modulation current $\delta I_e$ ~20 nA$_{RMS}$ is applied by adding an AC modulation $\delta V_b$ to the DC level $V_b$ through a unity-gain isolation amplifier (Burr Brown ISO124P) and 100× voltage divider; the output is independent of frequency between 4 and 1000 Hz and exhibits noise of ~$10^{-6}$ $V/\sqrt{Hz}$. The typical modulation frequencies for our measurements range between 7 - 11 Hz.

For phonon detection, the detector is voltage biased in the subgap regime ($V_d \sim \Delta_d/e$) with the Josephson current suppressed. The detector signal comprises a steady state plus a modulated component, as indicated in figure 8a. The steady state DC detector current $I_d = $ ~1 to 2.5 nA for emitter voltages $V_e = 0.35$ to 5 mV as shown in figure 8b. For DC detector tunnel currents $I_d$ up to 1.5 times the unperturbed (thermal) level of the steady state detector current, we treat $\tau_{rec}$ as being constant and therefore equation (4) as being valid and {Eff} being fixed [41, 53]. We checked this assumption by raising the device temperature until $I_d$ rose by a factor of 3, and observed very small change in the differential transfer function $\delta I_d/\delta I_e$. Thus for $I_d < 1.5$ times its thermal level, we can safely assume that the detector response remains linear with incident phonon flux. (We note that for $I_d >$ 1.5 times its thermal level, the detector response may be nonlinear with the incident phonon flux.) In our devices $\tau_{rec}$ may be limited by magnetic flux trapped in the Al detector film as well as by quasiparticle population [41].

The modulated AC detector current (also differential response or differential transfer function) of our detector (figure 8c), which represents the modulated portion of the incident phonons, is isolated via a low-noise current pre-amplifier (DL 1211) and a lock-in amplifier (SRS 830) over a range from 0 to ~1 pA$_{RMS}$. As shown in figure 8c, the emitter tunnel junction turns on at emitter voltage above $V_e = 2\Delta_e/e$: the step in detector response at $V_e = (2\Delta_e + 2\Delta_d)/e$ occurs because the emitted relaxation phonons (peak energy = $eV_e - 2\Delta_e$) above this voltage are energetic enough to break Cooper pairs in the detector (gap energy $2\Delta_d$, i.e. ~90 GHz). When $V_e = (2\Delta_e + 4\Delta_d)/e$, we observe a further change in detected signal level, as the emitted relaxation phonons acquire enough energy to break multiple cooper pairs in the detector (See also equation (3)). We have also considered the effect of microwave Josephson radiation on the detector signal [29]. In one spectrometer, we biased the emitter at $V_e = 0$ and modulated the Josephson branch of the emitter I-V curve. We observed zero detector response. We conclude that our measurement is not influenced by Josephson radiation or inductive coupling of the emitter Josephson current into the detector.

The peak frequency of the emitted relaxation phonon distribution is related to the emitter bias voltage as $(eV_e - 2\Delta_e)/h$. The feature in figure 8c at $V_e \sim 4$ mV is believed to be due to backscattering by oxygen impurities in the silicon. This peak was observed at ~870 GHz in past studies of STJ phonon spectroscopy [54], [55]. While this behavior confirms that our aluminum STJ-based spectrometer emits a strong and tunable signal well above 800 GHz, we note that at such high frequencies (figure 2d), we estimate only ~20% of the total phonon power to be at the peak frequency of $(eV_e - 2\Delta_e)/h$.

In figure 8d, we present voltage-biased I-V curves of a detector recorded while varying the emitter voltage from 0 to ~5 mV. (We note that in this detector we were unable to suppress Josephson current below ~5 nA.) For emitter voltage $V_e = 0$ V, the subgap current at detector voltage $V_d = \Delta_d/e$ (180 µV) is exactly the same as that shown in figure 7c at a temperature of ~313 mK. As a larger and larger phonon flux is transmitted to the detector, the total quasiparticle density in the detector increases well beyond the thermal level, and the detector current rises. In figure 8e, we calculate the differential conductance (dI/dV) from the subgap I-V measurements of figure 8d. The conductance of the detector remains essentially the same as emitter voltage is varied. At the typical bias point of $V_d = \Delta_d/e$, conductance $G$ remains fixed at ~5×10$^{-6}$ / Ω. The only difference is in the total current level.

These measurements motivate a simplified equivalent circuit model for our STJ phonon detector, shown in figure 8f. The phonon detector is modeled as a current source in parallel with a resistance $1/G$. The DC current $I_d$ and modulated current $\delta I_d$ follow the incident flux of phonons. The detector is in series with the current amplifier (input impedance $R_{AMP}$ and current through $I_{AMP}$) and line resistance $R_{LINE}$. The bias point on the detector is maintained by an isolated voltage source (Stanford Research SIM928, output through a 10$^5$ voltage divider) across the entire network. Typical values for $R_{LINE}$ and $R_{AMP}$ are ~70 Ω and 2 kΩ respectively ($R_{AMP}$ is the manufacturer's specification). This model, and the measurements of figures 8d and 8e, makes clear that the STJ maintains a steady bias throughout our measurement range—even if $I_d$ rises by 1 nA, the bias across the STJ will change by only a few µV. Similarly, the current through the amplifier, accurately registers the modulated current $\delta I_d$ through the detector. Modulated amplifier current $\delta I_{AMP}$ equals $\delta I_d/(1 + G \cdot (R_{LINE} + R_{AMP}))$, which is only ~1% differerent than $\delta I_d$ for typical values of $R_{LINE}$, $R_{AMP}$ and $G$.

5. **Results of phonon spectroscopy measurements**

*5.1. Energy resolution and sensitivity*

The energy resolution of our measurement is limited by noise, by the band gap inhomogeneity of the emitter STJ, and by the modulation amplitude. Voltage noise across the emitter STJ adds random fluctuations to bias voltage $V_e$, while inhomogeneity in the emitter gap $\Delta_e$ likewise reduces precision of phonon energies. In practice, we assess these effects based on the width of the gap rise in the emitter I-V curve (figure 6c), typically ~60-80 µeV. The modulation current $\delta I_e$ applied to the emitter may also reduce energy resolution by adding a voltage oscillation of peak amplitude $2\sqrt{2}R_n\delta I_e$ to the emitter voltage $V_e = I_eR_n$. For typical emitter junction resistance $R_n$ ~ 800 Ω and $\delta I_e$~ 20 nA$_{RMS}$, this modulation envelope is only ~ 40 µeV, and therefore the bandgap inhomogeneity imposes the limit on energy resolution: ~60-80 µeV—corresponding to a frequency resolution ~15-20 GHz.

The sensitivity of the measurement is limited by detector noise, which may comprise electrical pick up noise, vibrational pickup in wiring and amplifier noise, as well as fundamental contributions such as Johnson noise in wiring and shot noise in the tunnel junction. Figure 9 shows a typical noise spectrum of the detector, exhibiting peaks in the spectrum at 60 Hz and its multiples due to power-line noise pickup, as well as an unexplained resonance at ~600 Hz. Wiring and apparatus to minimize noise are discussed in the section on instrumentation. Based on detector noise spectra such as figure 9, we typically choose modulation frequencies between 3 and 12 Hz, adding line-frequency notch filters and low-pass filters at the input of the preamplifier and lock-in amplifier to avoid amplifier overload. The lowest noise level obtained at modulation frequency of 11 Hz was ~60 $fA/\sqrt{Hz}$. We note that a tunnel junction passing a DC current of 1 nA should exhibit a shot noise of ~18 $fA/\sqrt{Hz}$ (assuming a Fano factor of 1), so our experimental noise is not far above the shot noise level. To reduce uncertainty in a spectral measurement, we typically repeat it 25 times and average the results. Considering the typical detector efficiencies {Eff}~0.1 (equation (4)) as well as acoustic-transmission and absorption factors $T_{SiAl}$ and $\alpha_{abs}$ (see equation (3)), we estimate the noise equivalent power (NEP) for phonon detection to be ~$10^{-15}$ $W/\sqrt{Hz}$, or ~2 x $10^7$ phonons of energy ~$2\Delta_d$ per second per $\sqrt{Hz}$. A comparative analysis of similar low temperature thermal detectors found similar sensitivities [56].

*5.2. Ballistic phonon propagation*

The ballistic nature of phonon transport is evidenced by comparing the differential detector response ($\delta I_d/\delta I_e$) of spectrometers with varying mesa widths, detector finger widths, blocked ballistic path, and offset line-of-sight between emitters and detectors (figure 10a-d). To enable measurements made with different detectors to be compared equivalently, we divide each measured value of $\delta I_d/\delta I_e$ by {Eff} for that detector to obtain the phonon transmission signal. Following equations (3) and (4), we expect the resulting scaled value to equal $T_{SiAl} \cdot \alpha_{abs} \cdot \frac{2e\delta \dot{n}_{ph,d}}{\delta I_e}$ for $2\Delta_d \leq \hbar\omega < 4\Delta_d$, and $T_{SiAl} \cdot \alpha_{abs} \cdot \frac{2e\delta \dot{n}_{ph,d}}{\delta I_e} \cdot (\hbar\omega/2\Delta_d - 1)$ for $\hbar\omega \geq 4\Delta_d$. Since $T_{SiAl}$ and $\alpha_{abs}$ are expected to be roughly the same from one detector to another, we do not rescale the data for these factors. We note that the quasiparticle diffusion length $\sqrt{D \cdot \tau_{rec}}$ is of order 100 µm, so that phonons reflected from the bottom of the Si chip and striking the wiring leads far from the junction or the mesa may also contribute to a measured 'background' signal level that is also subject to the same efficiency {Eff} as the signal resulting from phonons striking the detector finger [39]. The rate of ballistic phonons striking the detector finger, as measured by the differential detector response, is proportional to

$$\frac{\delta I_e}{e} \int dA_e \int \frac{d\Omega_d}{\pi} T_{AlSi} \cdot \cos\theta \cdot T_{SiAl} \cdot A_{foc}(\theta,\phi), \tag{5}$$

where $A_e$ is the fraction of emitter STJ visible from the detector, $\cos\theta$ is a Lambert law phonon emission distribution, $A_{foc}(\theta,\phi)$ is the phonon focusing factor, $T_{AlSi}$ and $T_{SiAl}$ are acoustic transmission factors described previously, and $d\Omega_d$ is the solid angle subtended by the detector with respect to the emitter STJ [30], [57], [58], [59], [60]. Figure 10a shows the phonon transmission signal between emitter and detector formed on different widths of mesas (7 µm (blue) and 10 µm (red)) with 6 µm detector finger widths. As the mesa width increases from 7 to 10 µm, the solid angle $\Omega_d$ subtended by the detector with respect to the emitter decreases; hence, the differential detector signal decreases as expected. We further verified the ballistic phonon transmission by varying the width, $W_f$, of the detector fingers. For a 10 µm mesa, we show the phonon transmission signal for a 6 µm wide (red plot) and a 3 µm wide (magenta plot) detector finger (figure 10b). The wider finger will subtend a larger solid angle; hence, the detector signal is larger as expected for the 6 µm wide detector finger shown in figure 10b. In figure 10c, we blocked the ballistic path between the emitter and detector by etching a trench into the mesa. The mesa width and detector finger widths are 10 µm and 6 µm respectively for both the bulk (open circle) and trench (hatched circle). The latter measurement reveals a significant portion of the transmitted phonon signals that are due to backscattering from the bottom of the chip ('background signal'). The difference between the trench transmission and the transmission through the mesa represents the dynamic range of our measurements. In figure 10d, we compare the phonon transmission signals for emitters and detectors that have a straight line-of-sight along the mesa width (along the <110> crystal direction, solid green plot), with emitter and detectors offset with line-of-sight offset by ~ 50° (near to the <100> crystal direction, open green plot). A slightly higher detector signal level is observed for the offset geometry. For this geometry, the ballistic signal is affected by phonon focusing—the attenuation or enhancement of phonon propagation in preferred direction in an anisotropic crystal such as silicon [59]. In silicon crystals, the phonon focusing factor is ~2 times higher in the <100> direction than in the <110> direction [60]. These measurements evince the sensitivity of our phonon spectrometer to submicron variations in device geometry. We point out, however, that the measured differential response of the detector must be scaled by the efficiency factor {Eff} in order to compare measurements from different detectors. In figure 11, we replot the results in figure 10a (phonon transmission through different mesa widths) with the unscaled detector response $\delta I_d/\delta I_e$, and we show that with typical detector efficiency factors {Eff}~0.1, there is an order of magnitude difference between the scaled and unscaled signals.

## 6. Conclusion

We have designed and microfabricated a phonon spectrometer utilizing superconducting tunnel junction transducers for the emission and detection of hypersonic (100 to ~870 GHz) acoustic phonons in silicon microstructures. We model the phonon emission profile of the modulated STJ phonon spectrum considering the electron-phonon interactions within the superconductor films of the emitter STJ, and we also model the phonon detector behavior by considering quasiparticle-phonon interactions. Our energy resolution of ~60-80 µeV, corresponding to a frequency resolution of ~15-20 GHz, is about 20 times better than the energy resolution obtainable from conventional thermal transport measurements, which rely on a Planck distribution of phonons. We have demonstrated that with a phonon detection noise equivalent power, NEP, of $10^{-15}\ W/\sqrt{Hz}$, the sensitivity of our STJ phonon detectors is comparable to similar low temperature thermal detectors that are available. The design of our spectrometer—comprising a silicon mesa with STJs on the sides—serves as a good platform for phonon transport studies. The ballistic phonon transmission through the mesa alone can be distinguished from backscattering from the substrate by subtracting the mesa-with-trench phonon transmission signal from the mesa-without-trench signal – a method which eliminates the need for more complicated suspended structures as is typical for thermal conductance measurements. The silicon mesa platform is adaptable to studies of phonon transmission through nanostructures or nanomaterials by etching or depositing these into the ballistic path defined by the mesa. Finally, we have evinced spectrally resolved ballistic phonon transport in microstructures with submicron spatial resolution. Our STJ-based spectrometer provides a state-of-the-art tool for examining nanoscale effects on phonon transport.

**Acknowledgements**
The authors thank R.B. Van Dover, S. Baker and Cornell LASSP for loan of key equipment. We thank R.B. Van Dover, R. Pohl, and K. Schwab for helpful discussions and thank N.J. Yoshida, J. Chang, and A. Lin for help with apparatus and procedures. The work was supported in part by the National Science Foundation under Agreement No. DMR-1149036, and in part by the Cornell Center for Materials Research (CCMR) with funding from the Materials Research Science and Engineering Center program of the National Science Foundation (cooperative agreement DMR 1120296). M.A. was fully funded and O.O.O. was partially funded through support of the Energy Materials Center at Cornell (EMC[2]), an Energy Frontier Research Center funded by the U.S. Department of Energy, Office of Science, Office of Basic Energy Science under Award Number DE-SC0001086. This work was performed in part at the Cornell Nanoscale Facility, a member of the National Nanotechnology Infrastructure Network, which is supported by the National Science Foundation (Grant ECS-0335765). The authors declare no competing financial interests.

## Appendix A. Numerical example of phonon emission rate

As a numerical example, a typical aluminum STJ having normal-state tunnel resistance $R_n = 1000\ \Omega$ and biased at ~2.1 mV to produce a peak at 400 GHz, with a +/- 10 GHz modulation, will produce ~4 nW of total phonon power and about ~0.4nW of modulated phonon power. Because of the geometry of our spectrometer, only about 0.1% of this modulated power, or ~0.4 pW, will participate in the measurement. Of this about ~32%, i.e. roughly $5 \times 10^8$ phonons/sec, will be carried by the peak phonons in the 20 GHz band around 400 GHz; the remainder of the power (roughly 1 to $3 \times 10^9$ phonons/sec) is carried by phonons of energy lower than the peak. In contrast, a thermal source peaked at 400 GHz and emitting the same experimental power (~0.4 nW) will emit a similar fraction (0.4 pW) in the proper direction to participate in the experiment, but only ~3% of this, or roughly $5 \times 10^7$ phonons/sec, will be carried by phonons within +/-10 GHz of the peak. Of the remaining power, roughly half will be carried by phonons of frequency >410 GHz and half by phonons of frequency <390 GHz.

**Appendix B. Estimating the detector efficiency**

The measured differential tunnel current $\delta I$ in our detector will be proportional to the change in nearby quasiparticle density $\delta N_{QP}$ [30, 39, 41]:

$$\delta I = \frac{1}{2eR_n} \frac{\delta N_{QP}}{N_0} \frac{eV + \Delta_d}{\sqrt{(eV+\Delta_d)^2 - \Delta_d^2}} \quad \text{(B.1)}$$

where $R_n$ is the normal-state tunneling resistance of the junction, $N_0$ is the normal density of states at the Fermi level ($1.75 \times 10^{10} \cdot \mu m^{-3} eV^{-1}$ in Al), and the last factor reduces to 1.15 at our detector bias voltage $V = \Delta_d/e$ [30]. Equation (3) of the main text presents the differential rate of quasiparticle generation $\dot{n}_{QP,ph}$ as a function of differential rate $\dot{n}_{ph,d}$ of phonons incident on the detector. From this $\dot{n}_{QP,ph}$, we can determine the differential change in quasiparticle density $\delta N_{QP}$ by the steady-state assumption that the rate of quasiparticles generated must balance all quasiparticle loss rates. The primary loss process comprises diffusion of the quasiparticles into the attached wiring leads, followed by recombination into Cooper pairs [40]. We will assume that the tunneling itself does not contribute significantly to quasiparticle loss. For quasiparticles diffusing into a volume $vol$, the recombination loss rate is [30, 41]

$$\dot{n}_{QP,rec} = -\delta N_{QP} \cdot vol / \tau_{rec} . \quad \text{(B.2)}$$

The recombination time $\tau_{rec}$ is strongly sensitive to the total quasiparticle density $N_{QP} = N_{QP,th} + N_{QP,DC} + \delta N_{QP}$, where $N_{QP,th}$ is the thermally-activated quasiparticle density, $N_{QP,DC}$ is the quasiparticle density due to the full rate of incident phonons and $\delta N_{QP}$ is due to modulated incident phonons. However, as long as $N_{QP,DC} + \delta N_{QP} \ll N_{QP,th}$, we may treat $\tau_{rec}$ as constant [53]. At a temperature of 0.3K, $\tau_{rec}$ is roughly 30 μs [30, 42, 43]. To check the dependence of detector response on $N_{QP}$, we repeated one of our spectral measurements at a temperature of 0.36 K, at which $I_d$ was 3 times its value at 0.3K. We found that the detector response was degraded by only ~10% compared to the 0.3 K measurements. Thus, we expect that restricting $I_d$ to only 1.5 times its unperturbed (thermal) value should maintain the condition $N_{QP,DC} + \delta N_{QP} \ll N_{QP,th}$, and therefore maintain a consistent detector sensitivity. We note that the ~10% reduction upon raising the temperature to 0.36K is less than what would be predicted by the theory of Rothwarf and Taylor[53], suggesting that in our devices $\tau_{rec}$ is less temperature-dependent than this theory. One possible explanation is that magnetic flux trapped in the Al detector film contributes to the quasiparticle recombination rate in our detectors [41]. In some cases, cycling our devices above $T_c$ resulted in variations of a few percent in the measured phonon transmission signal, which is consistent with the presence of detector efficiency variations due to trapped flux.

In considering $\dot{n}_{QP,rec}$, the volume $vol$ primarily comprises the wiring trace attached to the finger, so we have $vol \cong W_{tr} \cdot d_{tr} \cdot \sqrt{D \cdot \tau_{rec}}$, where $W_{tr}$ and $d_{tr}$ are respectively the average total width and thickness of the trace, which in our devices are respectively 3.2 $\mu$m and 530 to 580 nm, and $\sqrt{D \cdot \tau_{rec}}$ is the diffusion length of the quasiparticles. For diffusion constant $D$ = 20 cm²/s, this length is ~250 $\mu$m [29, 40]. Thus the recombination rate found from equation (B.2) is $\dot{n}_{QP,rec} = -\delta N_{QP} \cdot W_{tr} \cdot d_{tr} \cdot \sqrt{D/\tau_{rec}}$. In steady-state we take the total rate of change of quasiparticle density to be zero, thus $\dot{n}_{QP,ph} + \dot{n}_{QP,rec} = 0$, and we find

$$\delta N_{QP} = \dot{n}_{QP,ph} \cdot \sqrt{\frac{\tau_{rec}}{D}} \cdot \frac{1}{(W_{tr} \cdot d_{tr})} \tag{B.3}$$

Thus the tunnel current may be related to the rate of quasiparticle generation by incident phonons found from equation (3) of the main text:

$$\delta I = \sqrt{\frac{\tau_{rec}}{D}} \cdot \frac{1.15}{2eR_n N_0 W_{tr} d_{tr}} \cdot \dot{n}_{QP,ph} \tag{B.4}$$

From equation (B.4) we may define a nondimensional efficiency factor {Eff} for each detector as the ratio of measurable current $\delta I$ to charge production rate $e\dot{n}_{QP,ph}$:

$$\{\text{Eff}\} = \sqrt{\frac{\tau_{rec}}{D}} \cdot \frac{1.15}{2e^2 R_n N_0 W_{tr} d_{tr}} \tag{B.5}$$

We note that the relatively large magnitude of quasiparticle diffusion length $\sqrt{D \cdot \tau_{rec}}$ (of order 100 $\mu$m) means that phonons reflected from the bottom of the Si chip and striking the wiring leads very far from the junction or the mesa may generate quasiparticles that register as a tunneling current at the detector STJ, therefore contributing to the measured backscatter signal level. It is interesting to think about whether we could reduce or eliminate the measured background level (which represents a source of experimental uncertainty) by redesign of the detector or wiring traces. However, we note from equations (B.2) to (B.5) that changes in the wiring trace dimensions may not achieve this goal: if we reduce the width $W_{tr}$ of the wiring traces in order to diminish the intercepted phonon flux, we also reduce the volume $vol$ occupied by the quasiparticles and thereby increase the tunneling efficiency {Eff} for both the quasiparticles formed in the finger and those formed in the leads.

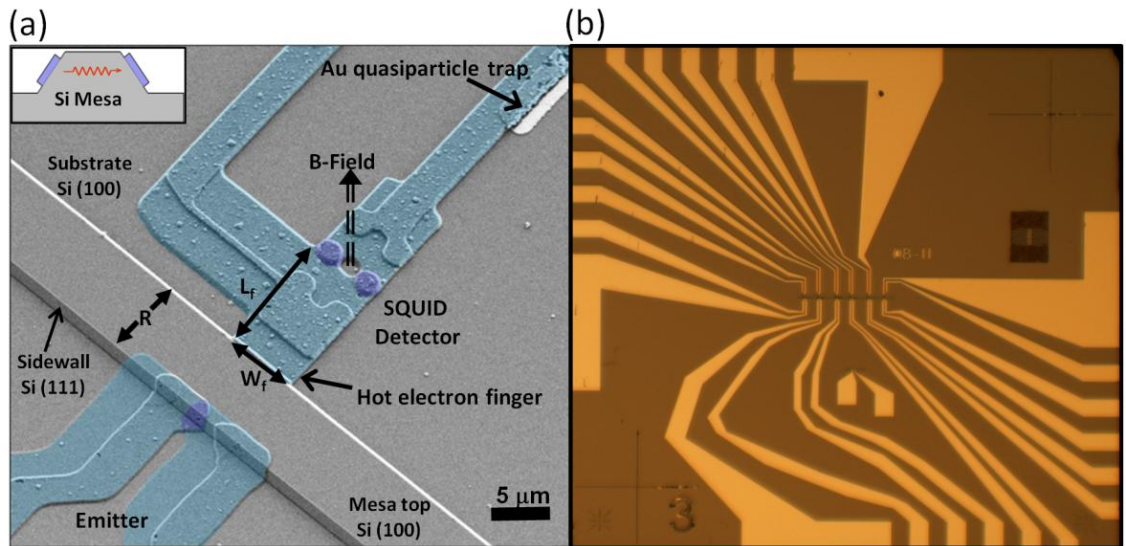

**Figure 1.** (a) False-colored SEM micrograph of completed phonon spectrometer. The STJ emitter is fabricated with the tunnel junction lying mostly on the sidewall of the 0.8 μm high mesa structure. The width of the mesa, R = 7, 10 or 15 μm. The mesa structure isolates a ballistic path for phonon transport between emitter and detector. The detector is fabricated in double-junction SQUID geometry with a hot electron finger for the collection of ballistically propagating phonons. Finger widths, $W_f$, were varied (1.5, 2, 3 or 6 μm) to observe the effects of geometry on phonon transmission. Magnetic field ~1 G is applied perpendicular to the SQUID detector for Josephson current suppression. 0.5 μm thick Silver film is deposited on the backside of the 500 μm thick silicon substrate to reduce phonon backscattering from the bottom of the substrate. (inset is a schematic of the side view of a silicon mesa with phonon transducers). (b) Optical microscope image of 4.5 square mm device comprising six spectrometers.

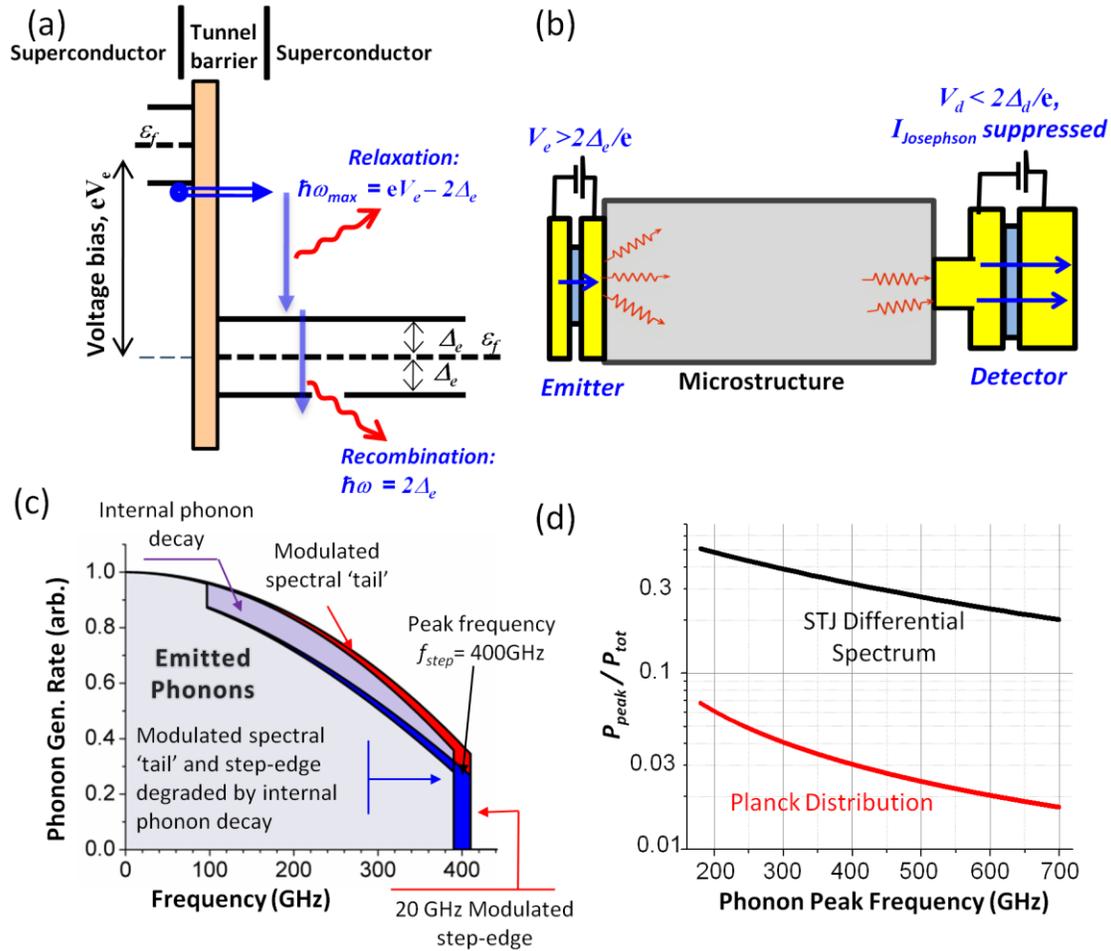

**Figure 2** (a) Energy diagram illustrating phonon emission with superconducting tunnel junctions. (b) Schematic diagram of the microfabricated phonon spectrometer. The diagram shows the emitter STJ on the left is biased above its superconducting gap ($2\Delta_e$) leading to phonon emission into the microstructure. The STJ detector on the right, which is biased below its superconducting gap and has its Josephson current suppressed, detects incident phonons with energy greater than $2\Delta_d$. The detector will capture incident phonons within the solid angle subtended by the detector finger with respect to the emitter. (c) Calculation of approximate phonon spectrum due to first-step relaxation of tunneled electrons, for a typical Al emitter STJ biased at ~2.1 mV to produce peak frequency $(eV_e - 2\Delta_e)/h = 400$ GHz and peak width $\delta\omega/2\pi = 20$ GHz. Differential portion of the spectrum produced by modulation is shaded red. Following phonon attenuation within the Al, remaining differential spectrum is shaded in dark blue (partially obscuring red-shaded area). (d) Calculation of $P_{peak}/P_{tot}$ for phonon spectrum emitted by quasiparticle relaxation in typical STJ emitter, found by integrating spectrum from part (c) and adding contributions of 2nd-step quasiparticle relaxation, quasiparticle recombination, and phonon reabsorption/reemission. For comparison, we calculate also $P_{peak}/P_{tot}$ of Planck distribution for a slice of width $\delta\omega/2\pi = 20$ GHz around the dominant phonon frequency.

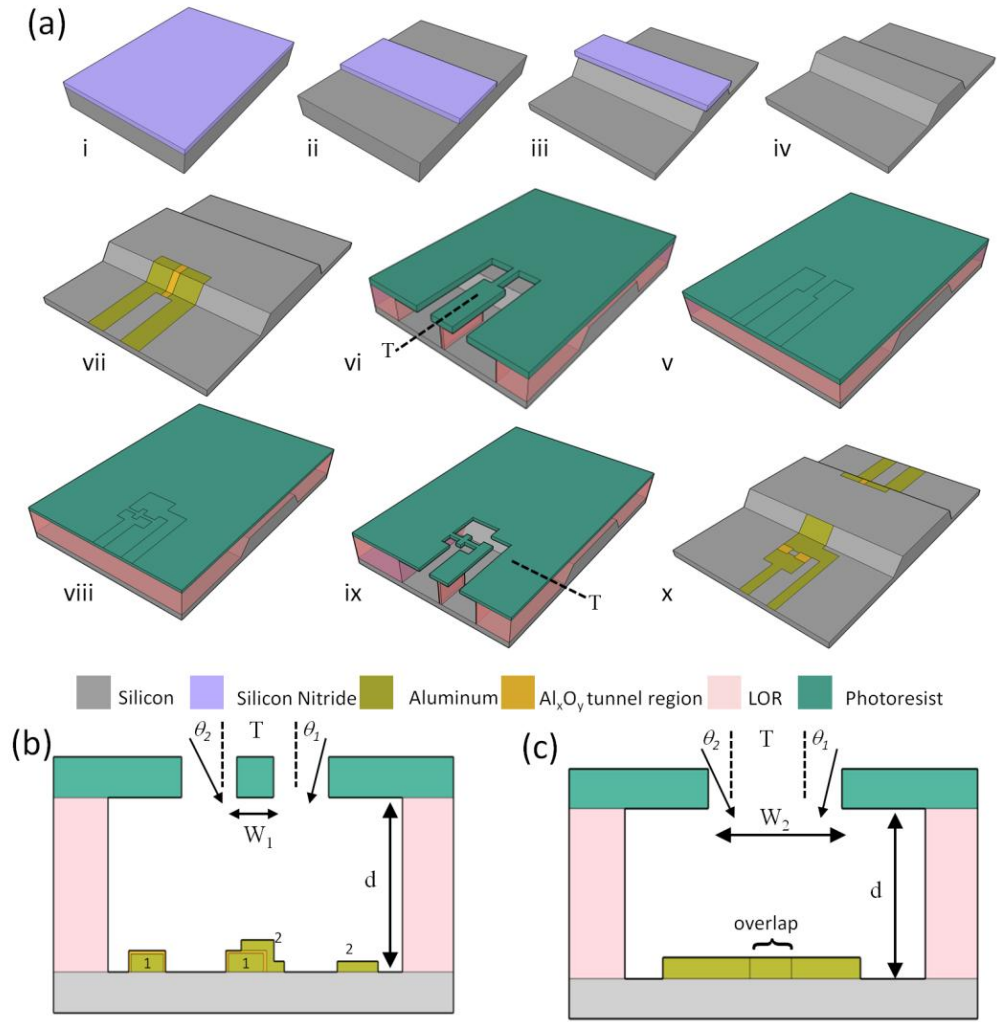

**Figure 3.** (a) Schematic of fabrication steps for STJ phonon transducers and mesas (steps i-x). (i) Low stress silicon nitride (~80 nm) grown on a silicon substrate (ii) Silicon nitride etched with $CHF_3/O_2$ to define mesa (iii) Anisotropic etching in KOH masked by $Si_xN_y$ (iv) Nitride etched in BOE (v) Bilayer of LOR and photoresist spun on mesas and emitter pattern and wiring trace is exposed on the top resist layer (vi) Exposed pattern developed, forming resist bridge across mesa sidewall (vii) Double angle evaporation and oxidation of aluminum followed by lift-off. Tilt axis (T) is indicated (viii) Bilayer of LOR and photoresist is re-spun and detector is patterned (ix) Detector pattern is developed forming Dolan bridges. Tilt axis (T) is indicated (x) Double angle evaporation and oxidation of detector film shown after lift-off. (b) Schematic diagram of double-angle evaporation. T indicates the tilt axis pointing into the page. $\theta_1$ and $\theta_2$ are the first and second deposition angles forming aluminum films numbered 1 and 2. First film is oxidized (orange colored region) before the second evaporation. The junction overlap width = $d \cdot (tan\theta_1 + tan\theta_2) - W_1$. (c) Schematic of overlapping films of the wiring traces with overlap width = $W_2 - d \cdot (tan\theta_1 + tan\theta_2)$. $W_2$ should be made wide enough to ensure overlap.

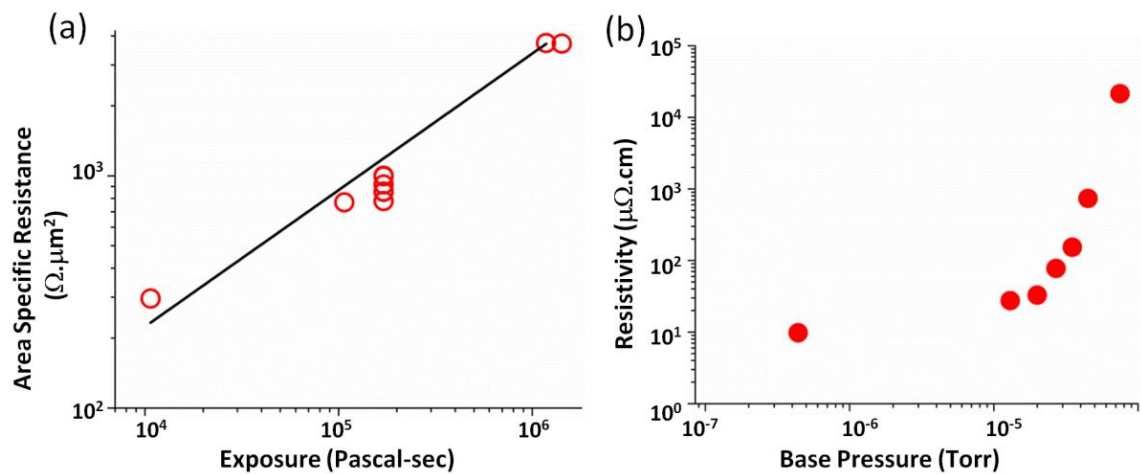

**Figure 4.** (a) Plot of Exposure parameter with the area specific resistance of 11 different tunnel junctions fabricated with 5 different exposure parameters. Power law fit ($f(x) = x^{0.59}$, adj-$R^2$ = 0.997). (b) Plots showing the effect of oxygen doping on Aluminum films. We measured the room temperature resistivity of aluminum evaporated in varying oxygen pressures starting with a film deposited at base pressure of 4.4x $10^{-7}$ Torr (No oxygen ).

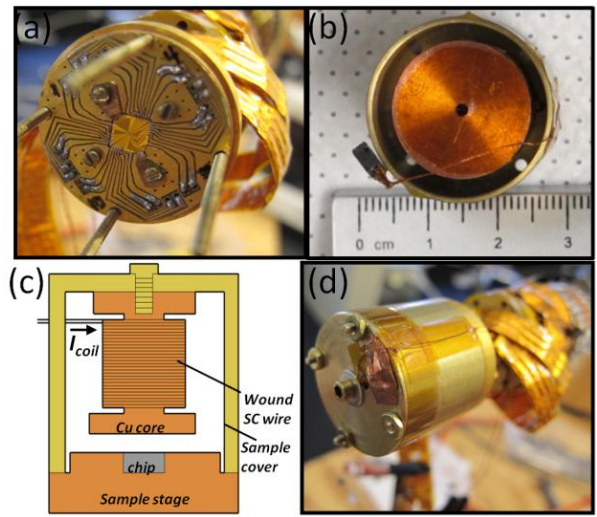

**Figure 5.** (a) Photograph of 4.5 sq. mm chip wire-bonded unto a gold plated sample stage of our He-3 cryostat. (b) Interior photograph of the sample box cover showing magnetic coil (~5000 turns of superconducting wire wound around a copper core). (c) Schematic of the coil assembly. Coil allows for the application of magnetic field perpendicular to the chip in order to suppress the Josephson current in the detector. Up to ~2.5 Gauss of magnetic field may be applied by passing current through the coil. (d) Sample cover mounted on the sample stage prior to cool down.

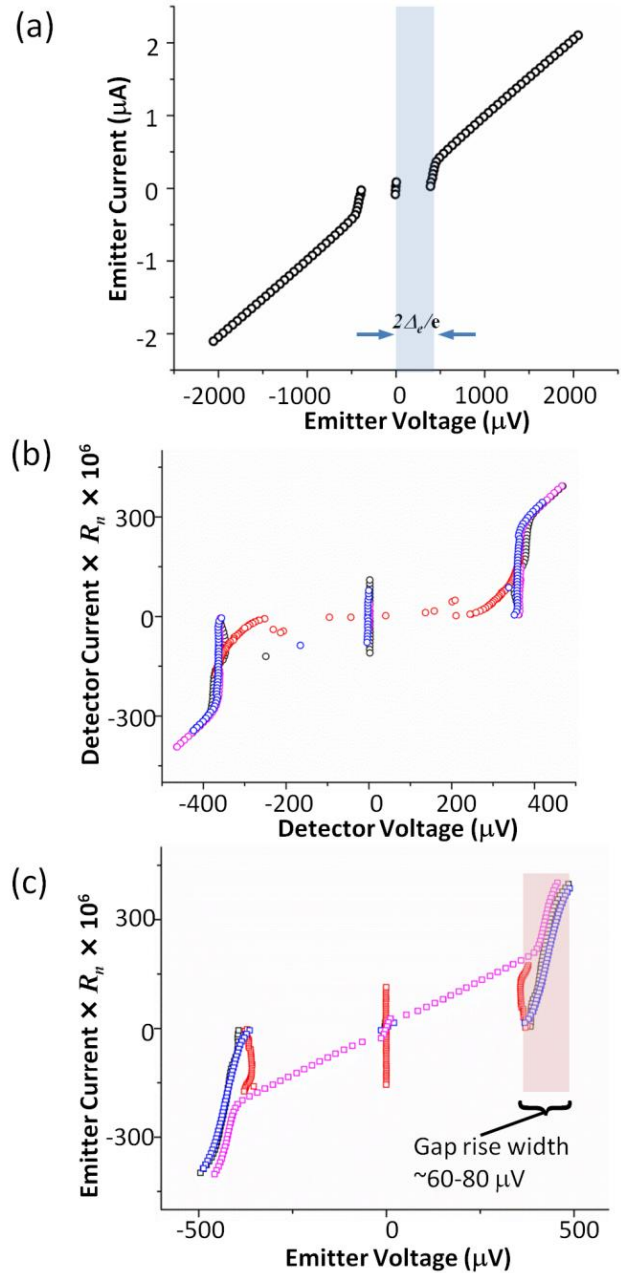

**Figure 6.** (a) Typical current biased I-V measurement of an emitter. The band gap of the junction $2\Delta_e$ is ~ 400 µeV. (b) $I \times R_n - V$ curves of four detectors taken without Josephson current suppression and focusing on the subgap regime. All the detectors are SQUIDs. Detector resistances range from 167 Ω (black), 213 Ω (magenta), 849 Ω (red), to 817 Ω (blue). Poorly filtered lines lead to rounding-off of gap rise as shown in the red plot. (c) $I \times R_n - V$ curves of four emitters focusing on the subgap region. Emitter resistances range from 935 Ω (black), 2250 Ω (magenta), 212 Ω (red), to 5556 Ω (blue). The plot illustrates how to identify common emitter problems: Magenta-colored plot shows a partly shorted device. The red plot exhibits severe 'back bending' of the gap rise due to overinjection and local suppression of superconducting gap, commonly seen in the case of low emitter resistance. The gap-rise width (black and blue plots) indicates inhomogeneity in superconducting gap $2\Delta_e$, which limits the energy resolution for phonon spectroscopy.

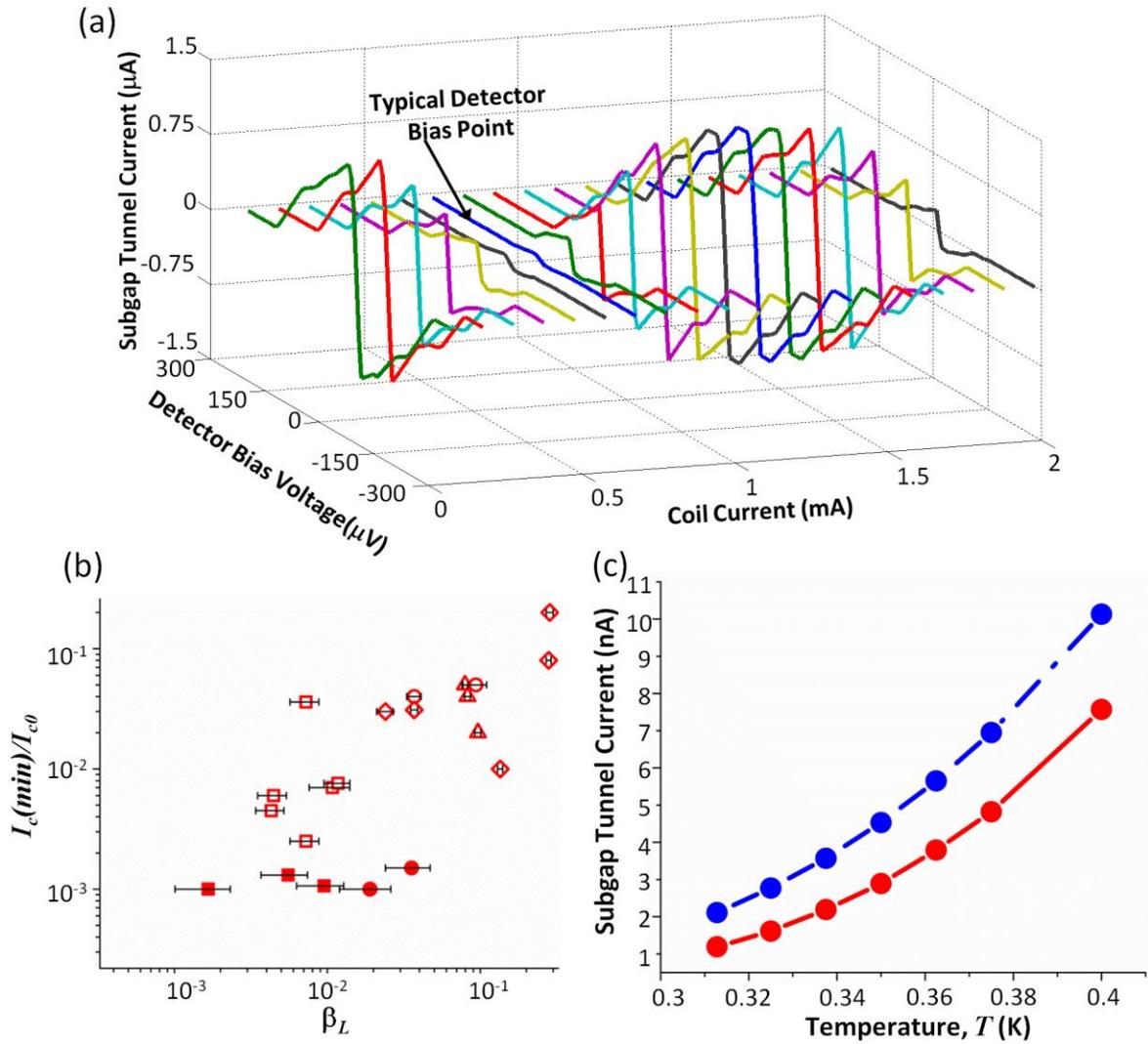

**Figure 7.** (a) Detector Josephson current suppression. The subgap tunnel current (before Josephson current suppression) is measured at each coil current from 0-2 mA as the bias voltage is swept from 300 µV to 300 µV. For operation in spectrometer, the detector is typically biased in the subgap region (~180 µV) and at external magnetic field ~1 Gauss (1.27 Gauss/mA) were the critical current is mostly suppressed. The plot also shows the periodic nature of the critical current with applied magnetic field. (b) Plot of $\beta_L (\frac{2LI_{c0}}{\Phi_0})$ versus the ratio of the minimum suppressed critical current to the calculated critical current $(I_c(min)/I_{c0})$ at T= 310 mK for several SQUID designs. Junctions formed on flat surface are represented by solid symbols, while the open symbols represent junctions formed on the sidewall; Loop areas vary from ≤ 2 µm² (squares), to ~10 µm² (circles), to ~120 µm² (triangles), and to ~180 µm² (diamond). (c) Temperature dependent subgap tunnel current (after Josephson current suppression) measured at $V_d = \Delta_d/e$ (180 µV) as function of temperature (red plot) and estimated tunnel current based on BCS prediction (blue plot).

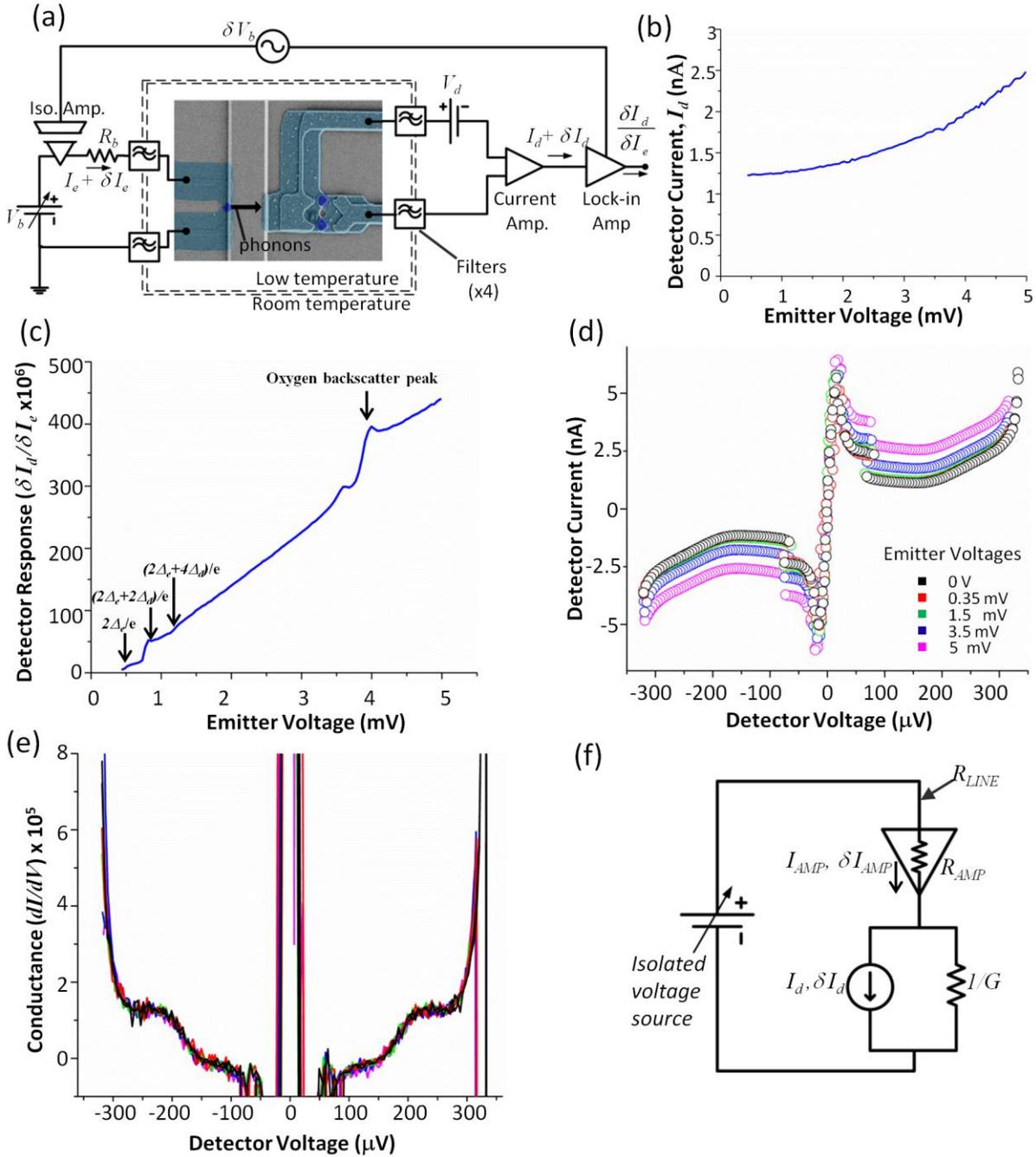

**Figure 8**. (a) Schematic of phonon spectroscopy measurements. (b) Steady state detector current, $I_d$. (c) Differential transfer function $\delta I_d/\delta I_e$, representing the fraction of emitted phonon flux that reaches the detector. The emitter tunnel junction turns on above $V_e = 2\Delta_e$ and emits detectable relaxation phonons only above $V_e = (2\Delta_e + 2\Delta_d)/e$. For $V_e = (2\Delta_e + 4\Delta_d)/e$, the emitted relaxation phonons may break multiple cooper pairs in the detector. The peak at ~4 mV represents resonant backscattering of oxygen impurities in the Si, typically seen at ~ 870 GHz [54]. (d) Voltage-biased detector I-V curves with varying emitter voltages and partially-suppressed Josephson current. (e) Differential conductance calculated from the I-V measurements in figure 8d (colors are same as in 8d). (f) Equivalent circuit model of detector as a current source. The DC current $I_d$ and modulated current $\delta I_d$ follow the incident flux of phonons.

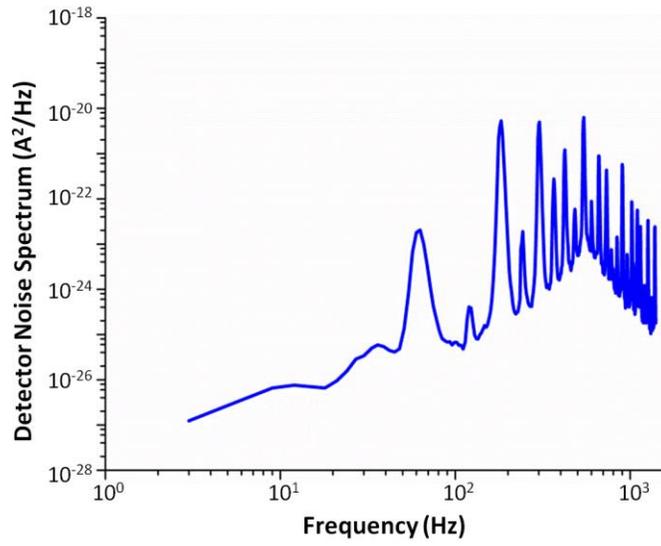

**Figure 9.** Detector noise spectrum. Modulation frequency for transport measurements range from 3 -12 Hz.

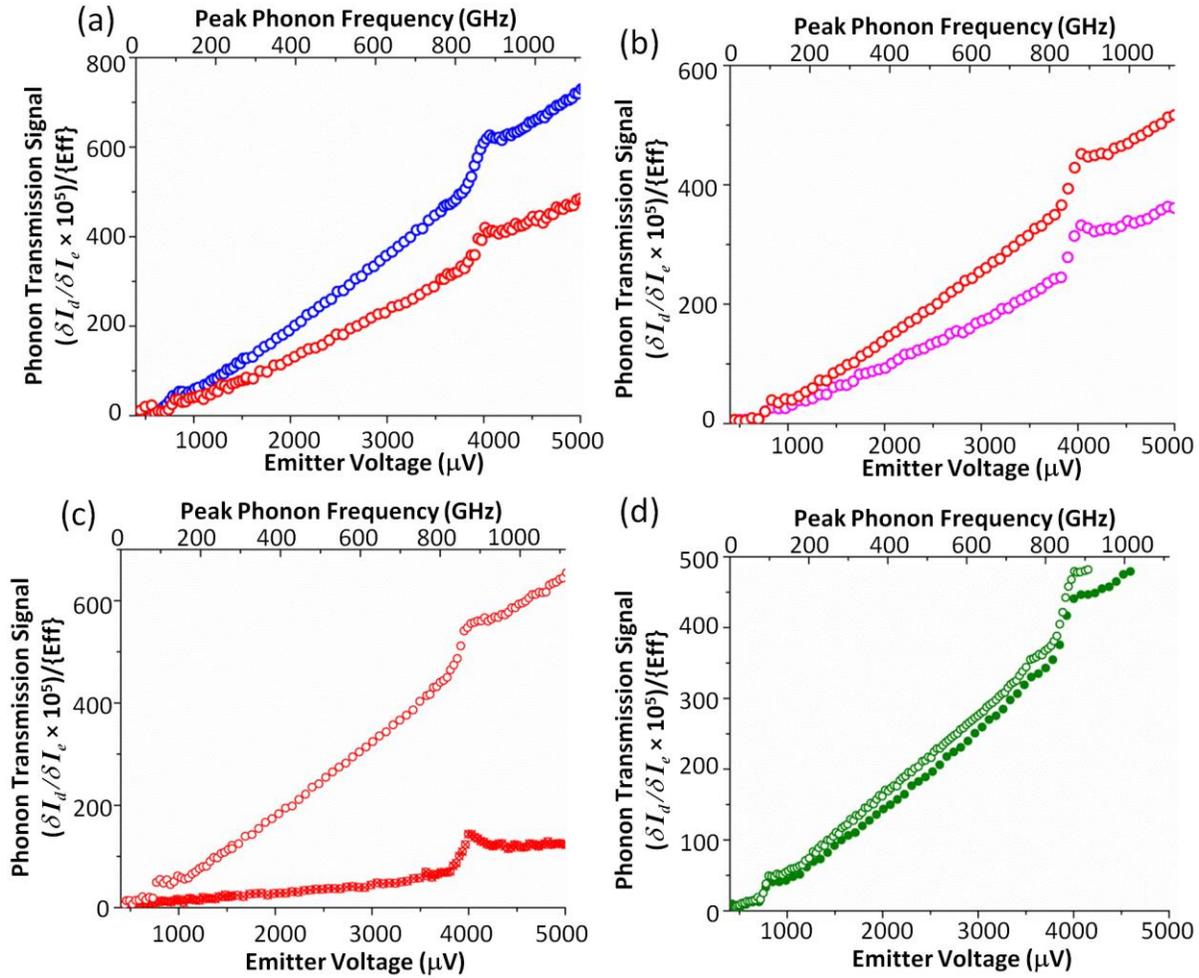

**Figure 10.** (a) Ballistic phonon transport measurements on different mesa widths. Detector signal level decreases as mesa width increases from 7 μm (blue) to 10 μm (red). The detector finger width is 6 μm in all cases. (b) Ballistic phonon transmission with varying detector finger widths. Detector signal collected by the 6 μm detector finger (red) is higher compared to the 3 μm detector finger (magenta). Mesa width is 10 μm in both cases. (c) Plots comparing phonon transmission through a mesa with (hatched red circle) and without (open red circle) a trench etched into the mesa. The trench blocks the line of sight between the emitter and detector. Mesa width is 10 μm and detector finger is 6 μm wide in both cases. (d) Ballistic phonon transport measurement with varying angle between emitter and detector. In the solid green circle plot, the emitter and detector have a straight line of sight, but in the open green circle plot the emitter and detector are offset by ~50 degrees. The mesa width is 7 μm and detector finger width is 3 μm in both cases. (Plots in figure 10 are not restricted to the region where the detector response is linear with incident phonon flux, i.e. for portions of the plot $I_d$ > 1.5 times its thermal level.)

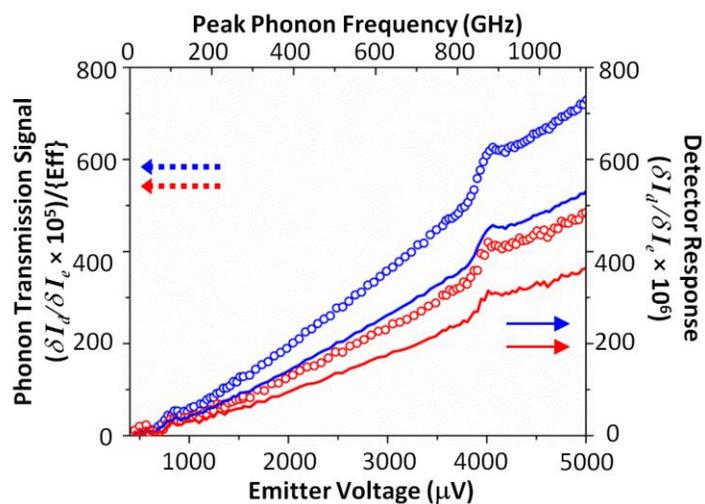

**Figure 11.** Comparison of the modulated detector response scaled by detector efficiency factor {Eff} (scatter plots and left axis) with the unscaled detector response (line plots and right axis) for plots in figure 10a. The measured detector response must be scaled by the efficiency factor in order to ascertain the phonon transmission signal. (Plots in figure 11 are not restricted to the region where the detector response is linear with incident phonon flux, i.e. for portions of the plot $I_d$ > 1.5 times its thermal level.)